\documentclass[aps,prb,reprint,amsmath,amssymb,floatfix]{revtex4-2}
\usepackage{graphicx}
\usepackage{bm}
\usepackage{physics}
\usepackage[colorlinks=true,citecolor=blue,linkcolor=blue,urlcolor=blue]{hyperref}
\usepackage{physics}
\usepackage{booktabs}
\usepackage{multirow}
\usepackage{epsfig,mathrsfs,color,latexsym,subfigure,marginnote,graphicx,verbatim,mathrsfs,color,array,amsmath,amsfonts,amssymb,graphicx}
\usepackage{hyperref}
\usepackage{dcolumn}
\usepackage{bm}

\usepackage[cal=euler]{mathalfa} 

\usepackage{lipsum}
\begin{document}

\title{Isolation of spin-valley locked nodal-line fermions in $d$-wave $\mathrm{AV_2X_2O}$ altermagnets}

\author{Pritesh Srivastava}
\affiliation{Department of Condensed Matter Physics and Materials Science, Tata Institute of Fundamental Research, Colaba, Mumbai 400005, India}

\author{Rahul Verma}
\affiliation{Department of Condensed Matter Physics and Materials Science, Tata Institute of Fundamental Research, Colaba, Mumbai 400005, India}

\author{Bahadur Singh}
\email{bahadur.singh@tifr.res.in}
\affiliation{Department of Condensed Matter Physics and Materials Science, Tata Institute of Fundamental Research, Colaba, Mumbai 400005, India}


\begin{abstract}
Crystalline symmetries stabilize topological states with distinct electronic properties, while altermagnets exhibit momentum-dependent spin splitting without net magnetization. Here, we combine first-principles calculations with a minimal tight-binding model to realize $C$-paired spin-valley-locked nodal-line fermions in the $d$-wave altermagnet $\mathrm{AV_2X_2O}$ (A = Rb, Cs, or K; X = Te, Se, or S). The low-energy electronic structure hosts coexisting spin-degenerate and spin-polarized nodal lines around $C_{4z}$-paired valleys near the Fermi level. The spin-polarized nodal lines are protected by the out-of-plane mirror symmetry $\mathcal{M}_z$ and remain robust against spin-orbit coupling. The minimal model reveals their microscopic origin and establishes a general design principle for their isolation. Layer engineering and electronic correlations serve as material-specific knobs for realizing these isolated spin-valley-locked nodal lines near the Fermi level. Our results establish the $\mathrm{AV_2X_2O}$ family as a versatile platform for exploring topological spin-valley locking in $d$-wave altermagnets.
\end{abstract}
\maketitle

\textit{Introduction}-- Interplay of crystalline symmetry and topology has fundamentally reshaped our understanding of quantum matter and yielded a rich spectrum of topological phases~\cite{Bansil2016colloquium,singh2023topology,Qi2011topological,bradlyn2017topological,po2017symmetry,hasan2021weyl,Lv2021experimental}. Crystalline symmetries protect band crossings from hybridization, stabilizing Dirac, Weyl, and nodal-line fermions with nontrivial surface states and unconventional transport. Among these, nodal-line semimetals, where band degeneracies extend along one-dimensional manifolds in momentum space, are particularly appealing. Their extended band crossings give rise to highly anisotropic electronic dispersions, drumhead surface states with a high density of states, and provide parent phases for topological insulators and Weyl semimetals upon symmetry breaking~\cite{burkov2011topological,PhysRevB.92.081201,Singh2018saddle,Singh2018spin}. Realizing nodal-line fermions with intrinsic spin-valley locking represents a new frontier where topology, spin, and valley become intrinsically intertwined, opening new opportunities for topological spintronics~\cite{Mak2014VHS}.

Spin-polarized electronic states are typically realized either through spin-orbit coupling (SOC) in inversion-symmetry-broken systems or through exchange splitting in ferromagnets~\cite{bychkov1984oscillatory,belopolski2019discovery}. In the former, broken inversion symmetry generates spin-momentum-locked states while preserving time-reversal symmetry ($\mathcal{T}$). In the latter, exchange splitting lifts spin degeneracy and, in suitable band structures, produces spin-polarized topological band crossings, as exemplified by the ferromagnetic semimetals $\mathrm{Co_3Sn_2S_2}$ and $\mathrm{Co_2MnGa}$~\cite{morali2019fermi,chang2016room,chang2017topological,sakai2018giant}. However, ferromagnets possess macroscopic magnetization and stray fields that hinder scalable spintronic applications~\cite{jungwirth2016antiferromagnetic}. Altermagnetism circumvents this limitation by combining zero net magnetization with crystal-symmetry-driven momentum-dependent spin splitting~\cite{vsmejkal2020crystal,vsmejkal2022emerging}. The resulting spin polarization reconstructs the low-energy electronic structure around symmetry-related ($C$-paired) valleys, creating opportunities for topological spin-valley locking with enhanced Berry-curvature responses and efficient spin-valley transport~\cite{Lai2025d,Hu2025catalog,Yar2026spin,jungwirth2026altermagnetic}. Recently, the layered $\mathrm{AV_2X_2O}$ family (A = Rb, Cs, or K; X = Te, Se, or S) has emerged as a promising class of $d$-wave altermagnets exhibiting large momentum-dependent spin splitting together with anomalous Hall-like responses, tunneling magnetoresistance, and spin Hall effects~\cite{ablimit2018v2te2o,jiang2025metallic,chang2026inverse,cui2023,PhysRevB.111.184437,zhang2025crystal,hu2026}. Recent ARPES measurements further revealed spin-degenerate and spin-valley-locked nodal lines around the $C_{4z}$-related valleys~\cite{hu2026}. However, the microscopic mechanism governing the emergence and isolation of the spin-valley-locked nodal-line fermions remains unexplored.

In this work, we realize isolated spin-valley-locked nodal-line fermions in the $d$-wave altermagnet $\mathrm{AV_2X_2O}$ through first-principles calculations and minimal tight-binding modeling. Their low-energy electronic structure hosts coexisting spin-degenerate and spin-polarized nodal lines around the $C$-paired valleys near the Fermi level. We show that the spin-polarized nodal lines are protected by $\mathcal{M}_z$ mirror symmetry and remain robust against SOC, whereas the remaining nodal crossings are selectively gapped. A minimal tight-binding model reveals their microscopic origin and establishes a design principle for realizing isolated spin-valley-locked nodal-line fermions. Guided by this principle, we demonstrate that layer engineering and electronic correlations provide practical routes to isolate these states at the Fermi level.

{\it Methods}-- First-principles calculations were performed within density-functional theory framework using the Vienna \textit{ab initio} Simulation Package (VASP)~\cite{VASP1996}. Exchange and correlation were treated using the Perdew-Burke-Ernzerhof generalized-gradient approximation~\cite{perdew1996generalized}, while electron correlations in the V $3d$ orbitals were described using the rotationally invariant DFT+$U$ method with $U_{\rm eff}=1$~eV~\cite{dudarev1998electron}. A plane-wave energy cutoff of 480~eV and a $\Gamma$-centered $12\times12\times6$ $k$-point mesh were employed for Brillouin-zone sampling. We relaxed lattice parameters and ionic positions until the residual forces were below $0.001$~eV/\AA. To characterize the topological features, a material-specific tight-binding Hamiltonian was constructed by projecting the Kohn-Sham states onto V $3d$, Te $5p$, and O $2p$ orbitals using the VASP2WANNIER interface~\cite{mostofi2008wannier90}. The resulting Hamiltonian was employed to identify nodal states and determine their symmetry representations using \textsc{WannierTools} and \textsc{IrRep}~\cite{wu2018wanniertools,iraola2022irrep}.

\begin{figure}
\centering
\includegraphics[width=0.48\textwidth]{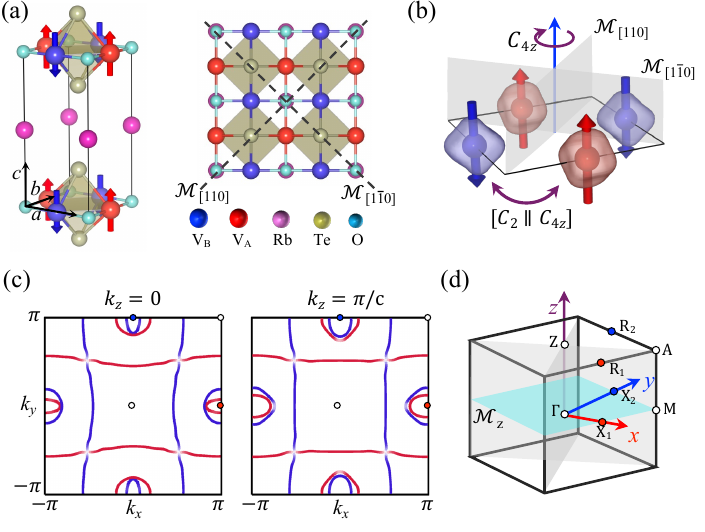}
\caption{Altermagnetic state of bulk $\mathrm{RbV_2Te_2O}$. (a) Crystal and magnetic structure. Red (blue) arrows denote the magnetic moments on the $V_A$ ($V_B$) sublattice. The two spin sublattices are related by the spin-lattice symmetry $[C_2||C_{4z}]$ and the diagonal mirrors $\mathcal{M}_{[110]}$ and $\mathcal{M}_{[1\bar{1}0]}$. (b) Real-space spin density around the V atoms. (c) Spin-resolved Fermi contours on the $k_z=0$ and $k_z=\pi/c$ planes. (d) Bulk Brillouin zone with the $C_{4z}$-paired valleys highlighted in red and blue. Gray planes denote the diagonal mirror planes relating the $X_1/R_1$ and $X_2/R_2$ valleys. Cyan plane denotes the horizontal mirror plane $\mathcal{M}_z$.}
\label{fig:CS}
\end{figure}

\textit{Altermagnetism-induced nodal-line fermions}-- We first describe the crystal and magnetic structure of bulk $\mathrm{RbV_2Te_2O}$ as a representative example of the $\mathrm{AV_2X_2O}$ family. It crystallizes in the tetragonal space group $P4/mmm$ (No.~123) and consists of $\mathrm{V_2Te_2O}$ layers separated by intercalated Rb atoms [Fig.~\ref{fig:CS}(a)]. The Rb atoms occupy the apical Wyckoff position $1b$ above the in-plane O sites ($1a$), while the V and O atoms form an inverse Lieb lattice within each layer~\cite{,chang2026inverse}. The magnetic ground state is a compensated collinear antiferromagnet with the Néel vector along the $[001]$ direction. Two inequivalent V sublattices, $\mathrm{V_A}$ and $\mathrm{V_B}$ carry opposite local moments of $\sim 2.1~\mu_{\mathrm B}$ per V atom. Their local V--O environments are related by a $\pi/2$ rotation, producing the characteristic $d$-wave altermagnetic order. Consequently, neither inversion nor $\mathcal T$ symmetry relates the two spin sublattices. Instead, the magnetic structure is invariant under the combined spin-lattice symmetry $[C_2\parallel C_{4z}]$ and preserves the diagonal mirror planes $\mathcal{M}_{[110]}$ and $\mathcal{M}_{[1\bar{1}0]}$, together with the out-of-plane mirror $\mathcal{M}_z$. Figure~\ref{fig:CS}(d) shows the bulk Brillouin zone with the $C_{4z}$-paired $X_{1,2}$ valleys on the $k_z=0$ plane and the $R_{1,2}$ valleys on the $k_z=\pi/c$ plane, as well as other high-symmetry momentum points and mirror planes.

\begin{figure}[b] 
\centering
\includegraphics[width=0.48\textwidth]{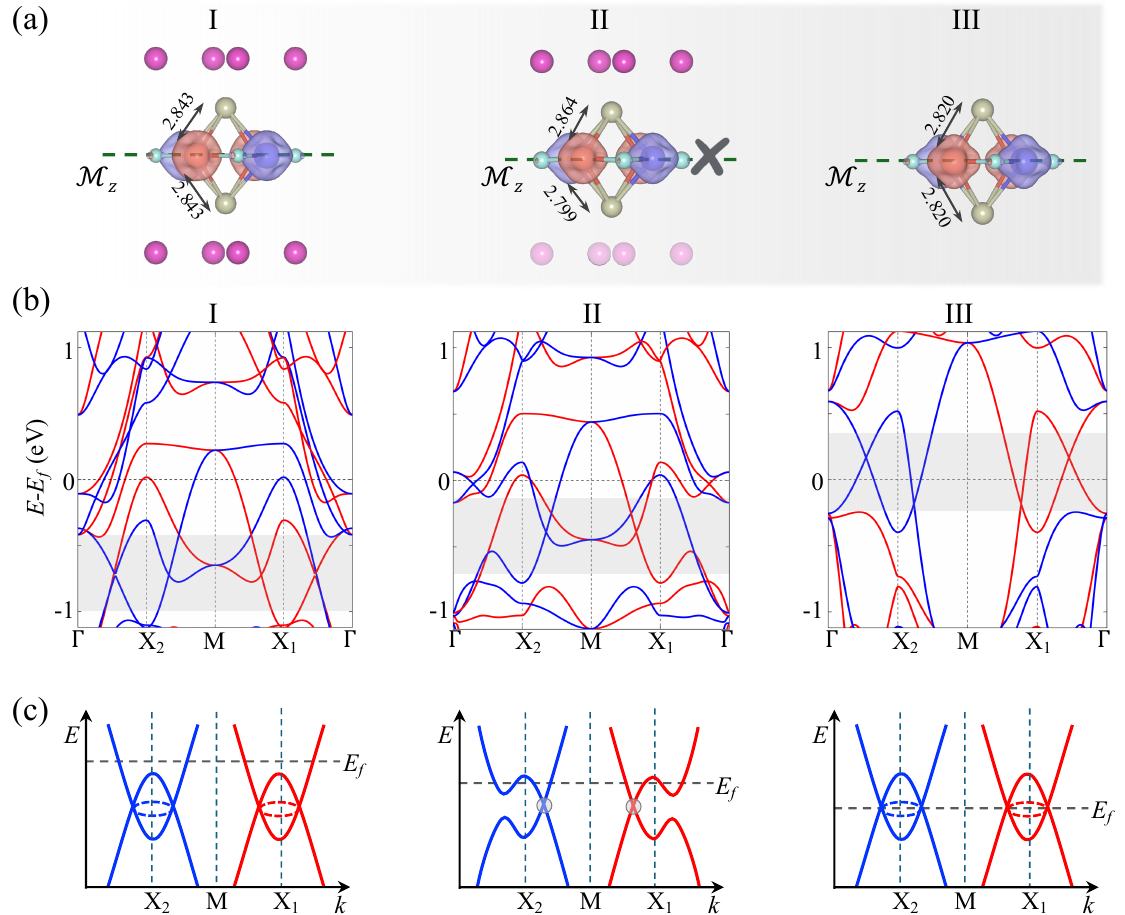}
\caption{Layer engineering of spin-valley-polarized nodal lines. (a) Crystal structures of the Rb-symmetric (I), Rb-asymmetric (II), and monolayer $\mathrm{V_2Te_2O}$ (III) with the corresponding spin densities. Structures I and III preserve $\mathcal{M}_z$, whereas II breaks it. (b) Corresponding electronic structures. Structure I hosts bulk-like nodal lines below the Fermi level, II gaps the mirror-protected nodal lines except along the Brillouin-zone boundary, and III isolates the spin-valley-polarized nodal line near the Fermi level. Highlighted regions indicate the relevant nodal crossings. (c) Evolution of the spin-valley-polarized nodal structure across the three layer geometries.}
\label{fig:tuning}
\end{figure}

Figure~\ref{fig:CS}(b) shows the real-space spin density of $\mathrm{RbV_2Te_2O}$. The spin density is predominantly localized on the V atoms and exhibits the four-lobed pattern characteristic of $d$-wave altermagnetism. The principal spin axes of the two sublattices are rotated by $\pi/2$, consistent with the underlying $[C_2\parallel C_{4z}]$ symmetry. The corresponding Fermi contours on the $k_z=0$ and $k_z=\pi/c$ planes are shown in Fig.~\ref{fig:CS}(c). They consist of spin-polarized pockets centered at the $X_{1,2}$ and $R_{1,2}$ valleys and open Fermi sheets spanning the Brillouin zone. Their spin polarization reverses across the Brillouin-zone diagonal, directly reflecting the real-space $d$-wave spin texture. The nearly identical Fermi contours on the $k_z=0$ and $k_z=\pi/c$ planes indicate weak out-of-plane dispersion and a quasi-two-dimensional electronic structure. This is further supported by the electron localization function (ELF) shown in Fig.~S1(b) of the Supplemental Material (SM). The ELF reveals strong covalent bonding within the $\mathrm{V_2Te_2O}$ layers and negligible charge localization in the interstitial region. Moreover, the nearly spherical ELF around the Rb atoms indicates weak hybridization with the $\mathrm{V_2Te_2O}$ layers, consistent with their role as electron donors. Accordingly, removing the Rb atoms leaves the low-energy band structure nearly unchanged and primarily shifts the Fermi level (see Fig.~\ref{fig:tuning}).

\begin{figure}[t!] 
\centering
\includegraphics[width=0.48\textwidth]{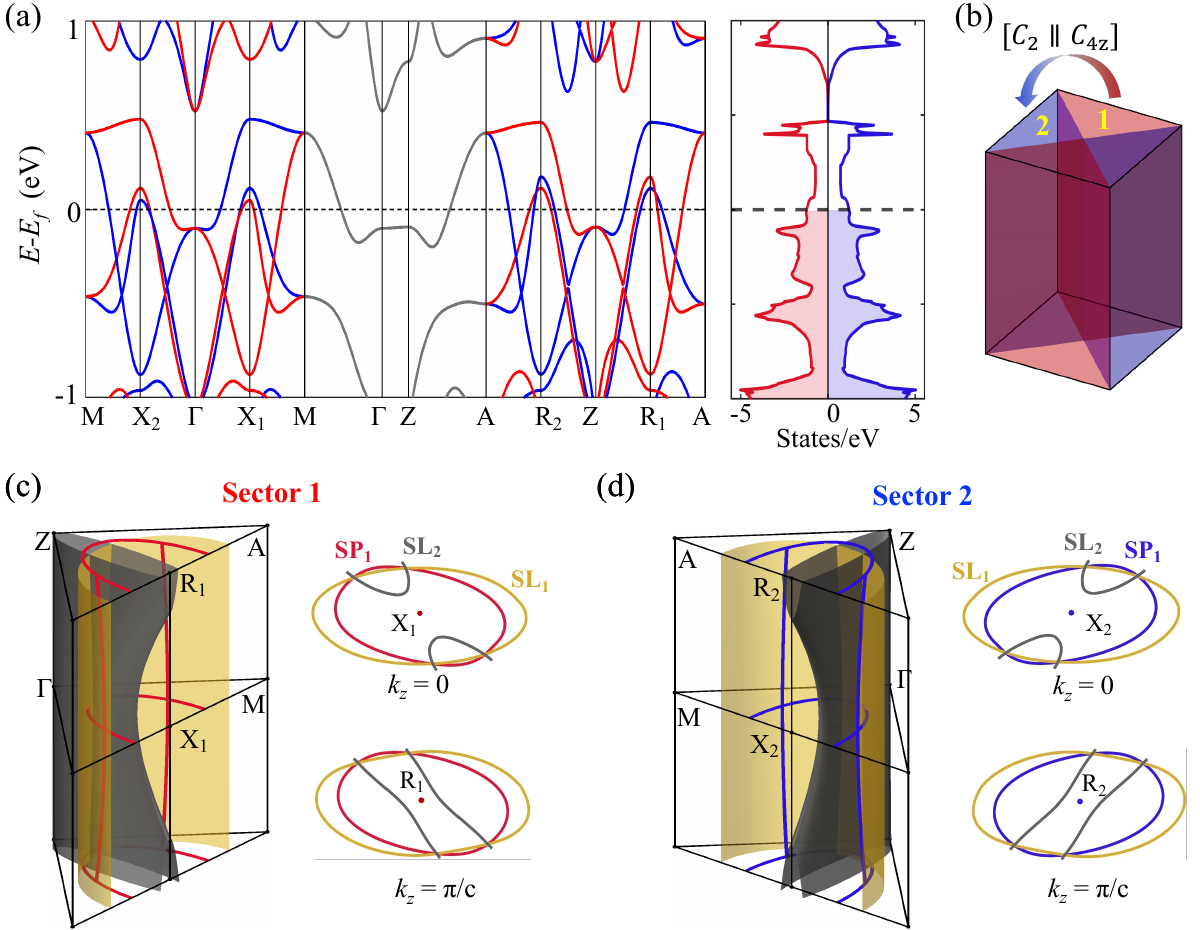}
\caption{Electronic structure and nodal configuration of bulk $\mathrm{RbV_2Te_2O}$. (a) Electronic structure without SOC. Spin-polarized band crossings appear near the Fermi level. (b) Spin-momentum locking partitions the Brillouin zone into spin sectors 1 and 2 related by the spin-lattice symmetry $[C_2\parallel C_{4z}]$. (c)-(d) Spin-polarized ($\mathrm{SP}_1$, red/blue) and spinless ($\mathrm{SL}_{1,2}$, gold/gray) nodal lines in the two spin sectors. Side panels show the nodal-line configuration on the $k_z=0$ and $k_z=\pi/c$ planes around the $X_{1/2}$ and $R_{1/2}$ valleys.}
\label{fig:bands}
\end{figure}

Figure~\ref{fig:bands}(a) shows the electronic structure of $\mathrm{RbV_2Te_2O}$ without SOC. The system is metallic, with two pairs of spin-polarized bands crossing near the Fermi level. Despite the momentum-dependent spin splitting, the spin-up and spin-down density of states remain identical, consistent with the compensated zero-net magnetization state. Without SOC, the Hamiltonian separates into two independent spin channels, $\mathcal{H}=\mathcal{H}^{\uparrow}\oplus\mathcal{H}^{\downarrow}$. The $[C_2\parallel C_{4z}]$ spin-lattice symmetry produces the characteristic $d$-wave altermagnetic spin splitting, $\Delta_{\rm AM}(\mathbf{k})=E^{\uparrow}(\mathbf{k})-E^{\downarrow}(\mathbf{k})\propto\mathbf{g}(\mathbf{k})\cdot\bm{\sigma}$, where $\mathbf{g}(\mathbf{k})=\alpha(k_x^2-k_y^2)\hat{z}$ is the effective nonrelativistic Zeeman field. Consequently, the spin splitting vanishes along the $k_x=\pm k_y$ planes and reaches its maximum near the $X_{1,2}$ and $R_{1,2}$ valleys, partitioning the Brillouin zone into two symmetry-related spin sectors, \textbf{1} and \textbf{2} [Fig.~\ref{fig:bands}(b)]. Related by the $[C_2\parallel C_{4z}]$ spin symmetry, these sectors exhibit identical band dispersions with opposite spin polarization. The four bands near the Fermi level comprise spin-split pairs of V $d_{xz}/d_{yz}$ and $d_{xy}$ orbitals from the two magnetic sublattices (See SM~\cite{Supplemental}). Crucially, these orbital manifolds exhibit markedly different spin splittings: the $d_{xy}$ bands split by $\sim0.06$~eV, whereas the $d_{xz}/d_{yz}$ bands split by $\sim1.35$~eV.

Figures~\ref{fig:bands}(c)-(d) reveal that these spin-resolved bands generate three distinct nodal structures. In sector \textbf{1}, the crossings between the $d_{yz}^\uparrow$ and $d_{xy}^{\uparrow/\downarrow}$ bands produce two nodal lines centered around the $X_1$ point in the $k_z=0$ plane: a spinless nodal line ($\mathrm{SL}_1$) and a spin-polarized nodal line ($\mathrm{SP}_1$). Owing to the quasi-two-dimensional electronic structure, both nodal lines exhibit negligible dispersion along $k_z$ and evolve into nearly cylindrical nodal surfaces extending across the Brillouin zone. The crossing between the opposite-spin $d_{xy}$ bands further generates a nodal surface ($\mathrm{SL}_2$) near the Fermi level. As $k_z$ increases, $\mathrm{SL}_2$ undergoes a Lifshitz transition from a closed to an open topology at $k_z\approx0.53\pi/c$ [Fig.~\ref{fig:bands}(e)]. Sector \textbf{2} hosts the identical nodal structures but with opposite spin polarization. Consequently, the spin-polarized nodal lines are valley selective, appearing around the $X_1$ and $R_1$ valleys in sector \textbf{1} and the $X_2$ and $R_2$ valleys in sector \textbf{2} [Fig.~\ref{fig:bands}(c)]. The opposite spin polarization of these symmetry-related nodal lines constitutes $C$-paired spin-valley nodal-line fermions in $\mathrm{RbV_2Te_2O}$.

\begin{figure}[b]
\centering
\includegraphics[width=\columnwidth]{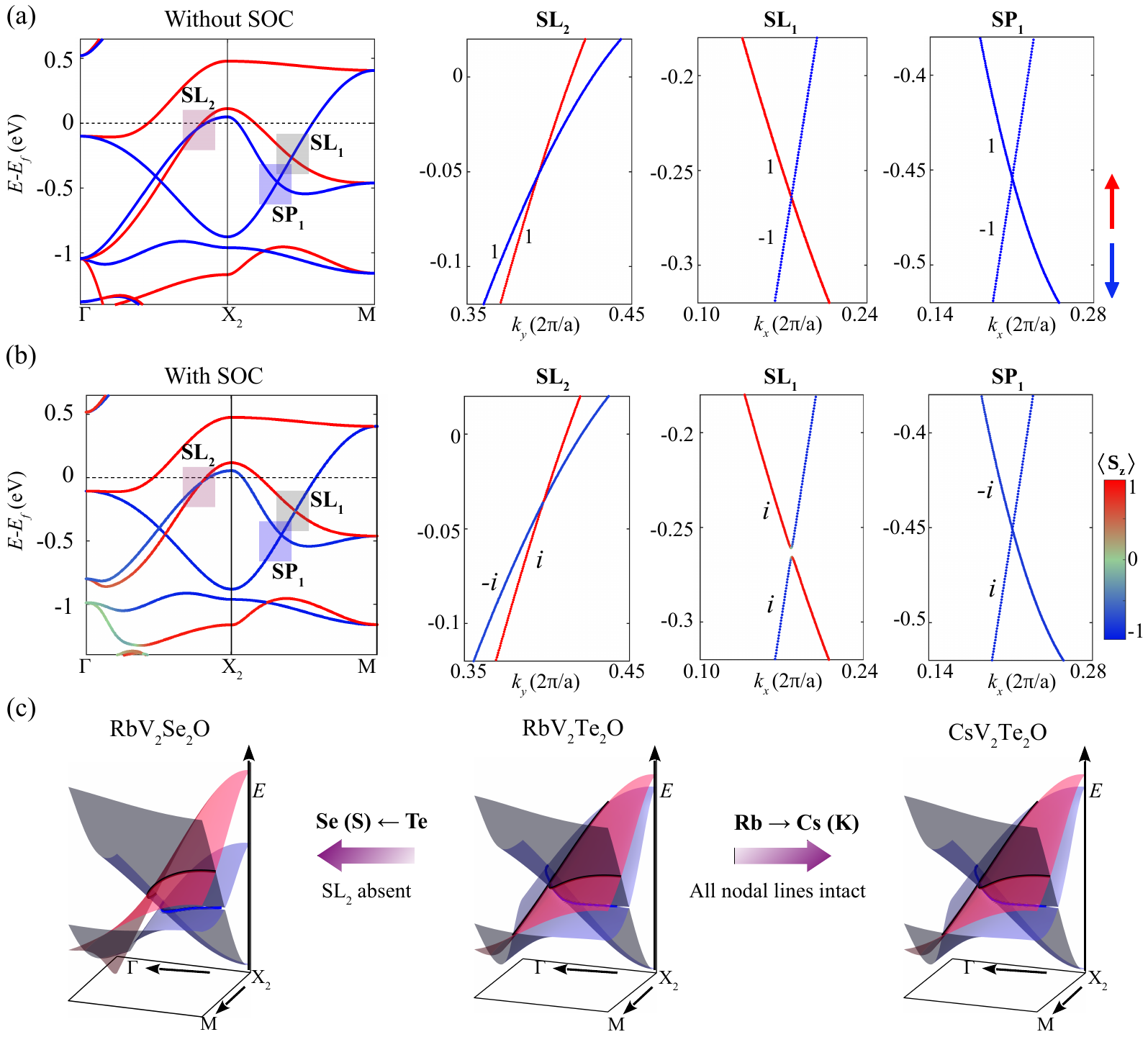}
\caption{Symmetry protection of nodal lines in $\mathrm{RbV_2Te_2O}$. (a) Electronic structure along the $X_2$ valley without SOC. Mirror eigenvalues and the local dispersion of the nodal crossings are shown in the side panels. (b) Electronic structure with SOC. The nodal crossings $\mathrm{SP}_1$, $\mathrm{SL}_1$, and $\mathrm{SL}_2$ are marked. (c) Electronic structure across the $\mathrm{AV_2X_2O}$ family. The nodal line $\mathrm{SL}_2$ is present only in $\mathrm{AV_2Te_2O}$ and disappears upon substituting Te with Se or S.}
\label{fig:SOC}
\end{figure}

In Fig.~\ref{fig:SOC}(a), we show the enlarged band dispersions around the $X_2$ valley in the $k_z=0$ plane, with all nodal lines marked. The nodal line $\mathrm{SL}_1$ exhibits a type-I conical dispersion, whereas $\mathrm{SL}_2$ is type II and originates from the crossing of $d_{xy}$ bands with opposite $\mathcal{S}_z$ eigenvalues. Without SOC, the Hamiltonian separates into two independent spin sectors, which prevents coupling between opposite-spin bands and protects both $\mathrm{SL}_1$ and $\mathrm{SL}_2$. In contrast, the spin-polarized nodal line $\mathrm{SP}_1$ is protected by the out-of-plane mirror symmetry $\mathcal{M}_z$. On the mirror-invariant planes ($k_z=0$ and $\pi/c$), $\mathcal{M}_z^2=1$, giving mirror eigenvalues $\pm1$. The two bands forming $\mathrm{SP}_1$ carry opposite mirror eigenvalues and therefore cannot hybridize. Away from these mirror planes, this protection is lost. Nevertheless, the quasi-two-dimensional electronic structure of $\mathrm{RbV_2Te_2O}$ limits the resulting gaps to below 1~meV.

Upon including SOC, the two spin sectors mix and spin-sector protection is uplifted. However, for the $[001]$ spin axis, $\mathcal{M}_z$ remains preserved and satisfies $\mathcal{M}_z^2=-1$, yielding the spinful mirror eigenvalues $\lambda=\pm i$. As shown in Fig.~\ref{fig:SOC}(b), the bands forming $\mathrm{SL}_2$ and $\mathrm{SP}_1$ carry opposite mirror eigenvalues and therefore remain $\mathcal{M}_z$ protected. In contrast, the bands forming $\mathrm{SL}_1$ share the same mirror eigenvalue, allowing a local gap to open. This nodal-line hierarchy remains almost intact throughout the $\mathrm{AV_2X_2O}$ family. Replacing Rb with K or Cs does not noticeably alter the $d_{xy}$ spin splitting, thereby preserving all nodal lines. In contrast, substituting Te with Se or S removes the type-II nodal surface $\mathrm{SL}_2$, whereas $\mathrm{SL}_1$ and $\mathrm{SP}_1$ remain intact (see SM). Nevertheless, the spinless nodal line remains close to the Fermi level above the spin-valley-locked nodal line, and their coexistence is a robust feature of the $\mathrm{AV_2X_2O}$ family.

\begin{figure}
\centering
\includegraphics[width=0.48\textwidth]{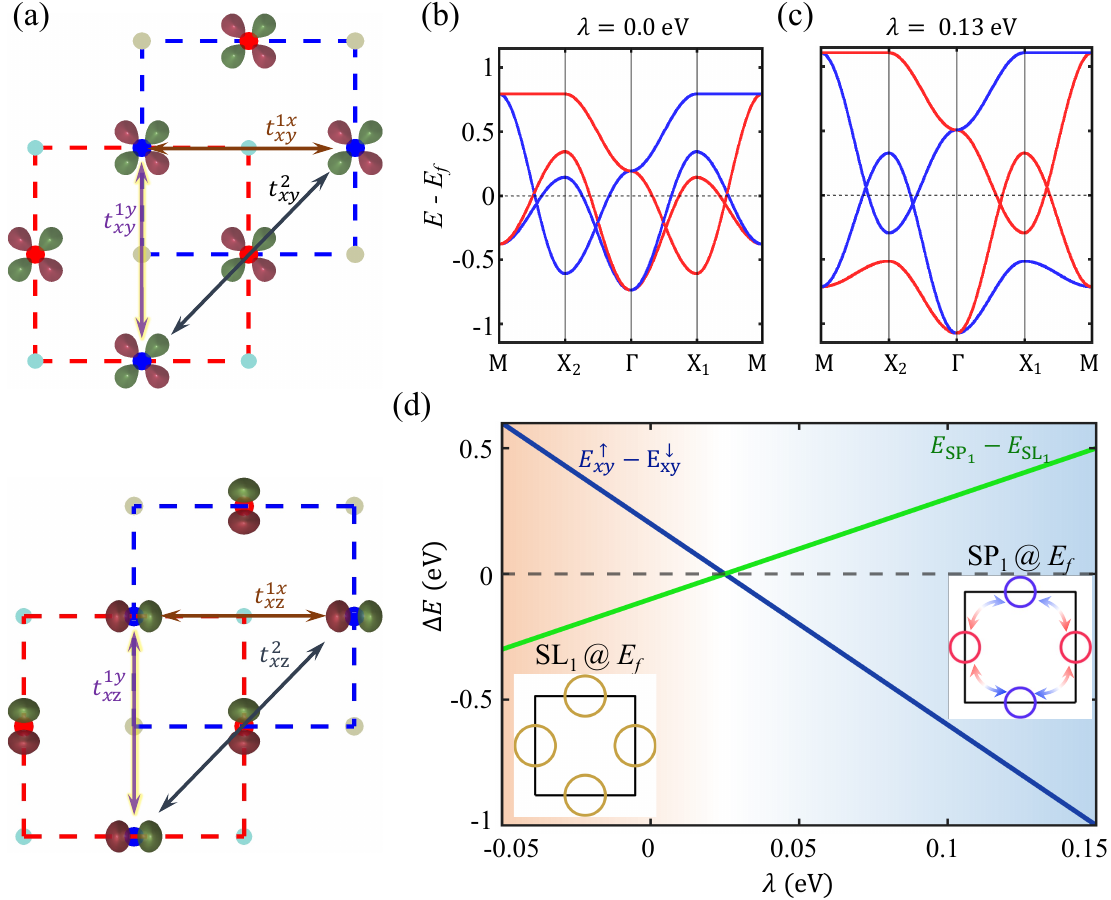}
\caption{Origin of isolated spin-valley-polarized nodal lines. (a) Inverse Lieb lattice of $\mathrm{RbV_2Te_2O}$. Top and bottom panels correspond to the $d_{xy}$ and $d_{xz/yz}$ orbitals, respectively. Hopping parameters are shown only for one magnetic sublattice. (b) Electronic structure for $\lambda=0$, corresponding to bulk $\mathrm{RbV_2Te_2O}$. (c) Electronic structure for $\lambda=0.13$ with isolated spin-valley-polarized nodal lines at the Fermi level. (d) Phase diagram of the $d_{xy}$ spin splitting and the nodal-line crossover to isolated spin-valley-polarized nodal lines as a function of $\lambda$.}
\label{fig:model}
\end{figure}

\textit{Microscopic origin of spin-valley polarized nodal lines}--
To elucidate the origin of the nodal-line structure, we construct a minimal tight-binding model for $\mathrm{AV_2X_2O}$. Guided by their quasi-two-dimensional electronic structure, we focus on the $k_z=0$ plane and employ a four-orbital basis, ${\bm \phi}=\{d_{yz}^{\uparrow},d_{xy}^\uparrow,d_{xz}^\downarrow,d_{xy}^\downarrow\}$, of the V atoms forming the inverse Lieb lattice~\cite{cheng2026realistic}. The spin-up and spin-down orbitals reside on the $\mathrm{V_A}$ and $\mathrm{V_B}$ sublattices, respectively. Without SOC, the $C$-paired spin sectors remain decoupled, while the same-spin sublattices are connected through the in-plane O and out-of-plane Te atoms [Fig.~\ref{fig:tuning}(a)]. This yields the block-diagonal Hamiltonian
\begin{equation}
\mathcal{H} =
\begin{pmatrix}
H_{V_A}^{\uparrow} & 0\\
0 & H_{V_B}^{\downarrow}
\end{pmatrix},
\end{equation}
where
\begin{equation}
H^{\uparrow}_{V_A}=
\begin{pmatrix}
H_{yz}^{\uparrow} & 0\\
0 & H_{xy}^{\uparrow}
\end{pmatrix},
\qquad
H^{\downarrow}_{V_B}=
\begin{pmatrix}
H_{xz}^{\downarrow} & 0\\
0 & H_{xy}^{\downarrow}
\end{pmatrix}.
\end{equation}
The spin-orbital-resolved components are
$H_{\phi_i}^{\uparrow}=\epsilon_{\phi_i}+2t_{\phi_i}^{1x}\cos k_x+2t_{\phi_i}^{1y}\cos k_y+4t_{\phi_i}^{2}\cos k_x\cos k_y$
and
$H_{\phi_i}^{\downarrow}=\epsilon_{\phi_i}+2t_{\phi_i}^{1x}\cos k_y+2t_{\phi_i}^{1y}\cos k_x+4t_{\phi_i}^{2}\cos k_x\cos k_y$,
where $\epsilon_{\phi_i}$ is the onsite energy associated with basis state $\phi_i$, $t_{\phi_i}^{1x}$ and $t_{\phi_i}^{1y}$ denote the directional nearest-neighbor V--V hoppings mediated by the in-plane V--O--V or out-of-plane V--Te--V pathways, while $t_{\phi_i}^{2}$ describes the next-nearest-neighbor hopping [Fig.~\ref{fig:model}(a)]. Details of various parameters for $\mathrm{AV_2X_2O}$ are given in Fig.~S3~\cite{Supplemental}. 

The minimal model identifies the asymmetry between the in-plane V--O and out-of-plane V--Te hopping pathways connecting the magnetic sublattices as the key microscopic mechanism controlling the nodal-line fermions. This asymmetry controls the spin density around each magnetic sublattice and directly determines the valley $d_{xy}$ spin splitting, thereby fixing the position of the spin-valley-locked nodal line relative to the Fermi level. To quantify this effect, we introduce an anisotropy parameter $\lambda$ that modifies the hopping amplitudes according to $t^{1y}_{xy}(\lambda)=t^{1y}_{xy}+\lambda$ and $t^{1x}_{xy}(\lambda)=t^{1x}_{xy}-\lambda$, and the onsite energy as $\epsilon_{xy}(\lambda)=(1+c\lambda)\epsilon_{xy}$, where $c$ is a material-dependent scaling constant. For $\lambda=0$, the $d_{xy}^{\uparrow}$ band lies above (below) the $d_{xy}^{\downarrow}$ band at the $X_2$ ($X_1$) valley [Fig.~\ref{fig:model}(b)]. As $\lambda$ increases, the $d_{xy}$ spin splitting decreases and changes sign at the critical value $\lambda_c=0.025$, inverting the ordering of the spin-split $d_{xy}$ bands. Consequently, the spin-valley-locked nodal line shifts toward the Fermi level and becomes energetically isolated from the spin-degenerate nodal lines [Fig.~\ref{fig:model}(c)]. The phase diagram in Fig.~\ref{fig:model}(d) demonstrates that $\lambda$ is the primary structural parameter for realizing isolated spin-valley-locked nodal lines.

The proposed mechanism can be realized in $\mathrm{AV_2X_2O}$ through (i) dimensional confinement and layer engineering by controlling the out-of-plane environment of the $\mathrm{V_2X_2O}$ slab and (ii) correlation-driven renormalization of the hopping amplitudes by increasing the Hubbard interaction $U$. Figure~\ref{fig:tuning} illustrates the former mechanism, while the SM demonstrates the latter. In the bulk-like structure [Fig.~\ref{fig:tuning}I], the $\mathrm{V_2Te_2O}$ slab is symmetrically sandwiched between two Rb layers, preserving the mirror symmetry $\mathcal{M}_z$. The relative strength of the in-plane V--O and out-of-plane V--Te hoppings gives rise to a nearly isotropic spin density around the magnetic sublattices, thereby preserving the bulk nodal structure. Charge transfer from the Rb layers, however, shifts the nodal structure below the Fermi level, as highlighted in Fig.~\ref{fig:tuning}(b). Removing one Rb layer [Fig.~\ref{fig:tuning}II] creates an asymmetric out-of-plane environment with inequivalent V--Te bonds. This enhances the spin-density anisotropy and reverses the valley $d_{xy}$ spin splitting. However, the broken $\mathcal{M}_z$ mirror symmetry gaps the mirror-protected nodal lines except along the Brillouin-zone boundary, yielding two-dimensional Weyl nodes. Removing both Rb layers [Fig.~\ref{fig:tuning}III] restores $\mathcal{M}_z$ in the freestanding $\mathrm{V_2Te_2O}$ monolayer. The shortened V--O bond strengthens the in-plane hybridization and further enhances the spin-density anisotropy. This stabilizes the reversed valley $d_{xy}$ spin splitting, bringing the spin-valley-locked nodal line to the Fermi level.

\textit{Discussion}-- Our results demonstrate that the $d$-wave $\mathrm{AV_2X_2O}$ altermagnets host coexisting spin-degenerate and spin-valley-locked nodal lines near the Fermi level. Their coexistence arises from the interplay of altermagnetic spin splitting and $\mathcal{M}_z$ mirror symmetry, while their tunability originates from the asymmetry between the V--V hopping pathways mediated by the in-plane O and out-of-plane Te atoms. This asymmetry controls the spin splitting of the low-energy V $d_{xy}$ states and enables the isolation of the spin-valley-locked nodal lines at the Fermi level. Layer engineering and tuning electronic correlations via the Hubbard interaction $U$ provide practical routes to control this nodal-line topology. Together, these results establish a microscopic design principle for topological spin-valley locking in layered altermagnets.

Experimentally, the weak hybridization between the Rb atoms and the $\mathrm{V_2X_2O}$ layers makes our proposal directly accessible. The Rb atoms primarily act as electron donors, while the low-energy electronic states reside in the $\mathrm{V_2X_2O}$ layers~\cite{ablimit2018v2te2o,hu2026}. Consequently, surface Rb atoms can be desorbed or etched without significantly altering the electronic structure, which exposes the isolated spin-valley-locked nodal line at the Fermi level. Recent experiments support this picture by demonstrating extremely weak Rb-mediated interlayer coupling and a quasi-two-dimensional electronic structure~\cite{zhang2025crystal,hu2026}. Together with the availability of high-quality single crystals and exfoliable thin films, these materials provide an ideal platform to probe the spin-valley-locked nodal-line fermions using spin-resolved ARPES and anisotropic magnetotransport. Looking ahead, the isolated spin-valley-locked nodal-line fermions offer a platform for valley-selective topological transport, Berry-curvature-driven nonlinear responses, and electrically tunable spin-valley phenomena in layered altermagnets.

\textit{Acknowledgements}-- This work is supported by the Department of Atomic Energy, Government of India, under Project Identification Nos. RTI4013 and RTI4015.

\bibliography{ref}
\lipsum[0]
\end{document}